\documentclass[prl,nobalancelastpage,superscriptaddress,twocolumn,nofootinbib]{revtex4}

\usepackage{graphicx}
\usepackage{amsmath, amsxtra, amssymb, mathtools, nccmath, xspace}
\usepackage[utf8]{inputenc}
\usepackage{float}
\usepackage{placeins} 
\usepackage{xcolor} 
\usepackage{nicefrac} 
\usepackage{xspace}     

\def\II{\hbox{$1\hskip -1.2pt\vrule depth 0pt height 1.6ex width 0.7pt\vrule depth 0pt height 0.3pt width 0.12em$}}
\newcommand{\pargap}{\\[12pt]}

\newcommand{\bra}[1]{\langle #1|}
\newcommand{\ket}[1]{\left|#1\right\rangle}

\newcommand{\ham}{\mathcal{H}}

\newcommand{\Rb}{\ensuremath{^{87}{\rm Rb}}\xspace}

\begin{document}
\title{
Quantum Kibble-Zurek mechanism and critical dynamics on a programmable Rydberg simulator
}

\newcommand{\emptyaddress}{}
\newcommand{\Caltech}{Division of Physics, Mathematics and Astronomy, California Institute of Technology, Pasadena, CA 91125, USA}
\newcommand{\Harvard}{Department of Physics, Harvard University, Cambridge, MA 02138,
USA}
\newcommand{\MIT}{Department of Physics and Research Laboratory of Electronics, Massachusetts Institute of Technology,
Cambridge, MA 02139, USA}
\newcommand{\ITAMP}{ITAMP, Harvard-Smithsonian Center for Astrophysics, Cambridge, MA 02138, USA}
\newcommand{\LKB}{Laboratoire Kastler Brossel, ENS, CNRS, Sorbonne Universit\'e, Coll\`ege de France, Paris, France }
\newcommand{\IQOQI}{Institute for Quantum Optics and Quantum Information, Austrian Academy of Sciences \& Center for Quantum Physics, University of Innsbruck, Innsbruck A-6020, Austria}

\author{Alexander Keesling}
\address{\Harvard}

\author{Ahmed Omran}
\address{\Harvard}

\author{Harry Levine}
\address{\Harvard}

\author{Hannes Bernien}
\address{\Harvard}

\author{Hannes Pichler}
\address{\Harvard}
\address{\ITAMP}

\author{Soonwon Choi}
\address{\Harvard}

\author{Rhine Samajdar}
\address{\Harvard}

\author{Sylvain Schwartz}
\address{\LKB}

\author{Pietro Silvi}
\address{\IQOQI}

\author{Subir Sachdev}
\address{\Harvard}

\author{Peter Zoller}
\address{\IQOQI}

\author{Manuel Endres}
\address{\Caltech}

\author{Markus Greiner}
\address{\Harvard}

\author{Vladan Vuleti{\'c}}
\address{\MIT}

\author{Mikhail D. Lukin}
\address{\Harvard}

\maketitle

\textbf{
Quantum phase transitions (QPTs) involve transformations between different states of matter that are driven by quantum fluctuations~\cite{Sachdev2009}. These fluctuations play a dominant role in the quantum critical region surrounding the transition point, where the dynamics are governed by the universal properties associated with the QPT. 
While time-dependent phenomena associated with classical, thermally driven phase transitions have been extensively studied in systems ranging from the early universe to Bose Einstein Condensates  ~\cite{Kibble1976,Zurek1985,delCampo2014,Navon2015}, understanding  critical real-time dynamics in isolated, non-equilibrium  quantum systems is an outstanding challenge~\cite{Polkovnikov2011}.  
Here, we use a Rydberg atom quantum simulator with programmable interactions to study the quantum critical dynamics associated with several distinct QPTs. By studying the growth of spatial correlations while crossing the QPT,  we  experimentally verify the quantum Kibble-Zurek mechanism (QKZM)~\cite{Polkovnikov2005, Zurek2005, Dziarmaga2005} for an Ising-type QPT, explore scaling universality, and observe corrections beyond QKZM predictions. This approach is subsequently used to measure the critical exponents associated with chiral clock models~\cite{Huse1982, Ostlund1981}, providing new insights into exotic systems that have not been understood previously, and opening the door for precision studies of critical phenomena, simulations of lattice gauge theories~\cite{Tagliacozzo2013,Weimer2010_b} and applications to quantum optimization~\cite{Farhi2000, Gardas2018}.}

\begin{figure}
	\includegraphics[width=\linewidth]{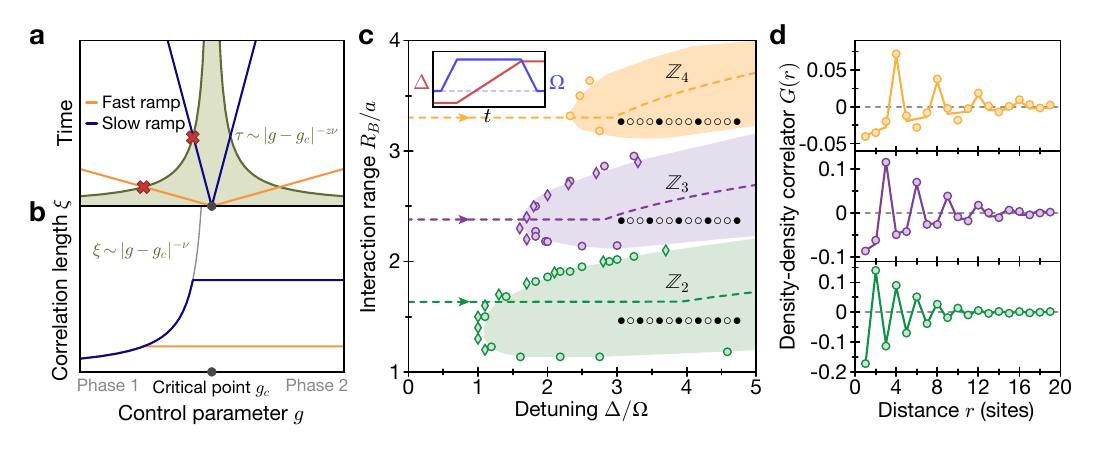}
	\caption{
	  \textbf{Quantum Kibble-Zurek mechanism (QKZM) and phase diagram.}
	\textbf{a}, Illustration of the QKZM. As the control parameter approaches its critical value, the response time, $\tau$, given by the inverse energy gap of the system, diverges. When the temporal distance to the critical point becomes equal to the response time, as marked by red crosses, the correlation length, \textbf{b,} stops growing due to nonadiabatic excitations.
    \textbf{c}, Numerically calculated ground-state phase diagram. Circles (diamonds) denote numerically obtained points along the phase boundaries calculated using (infinite-size) Density-Matrix Renormalization Group techniques (Methods). The shaded regions are a guide to the eye. Dashed lines show the experimental trajectories across the phase transitions determined by the pulse diagram shown as an inset. \textbf{d}, Measured (circles) density-density Rydberg correlations with fits to the expected ordered pattern (solid lines) consistent with $\mathbb{Z}_4$- (orange), $\mathbb{Z}_3$- (purple) and $\mathbb{Z}_2$-ordered (green) states. Error bars denote the standard error of the mean (s.e.m.) and are smaller than the marker size.
	}
	\label{fig:Phases}
\end{figure}

\begin{figure*}
	\includegraphics{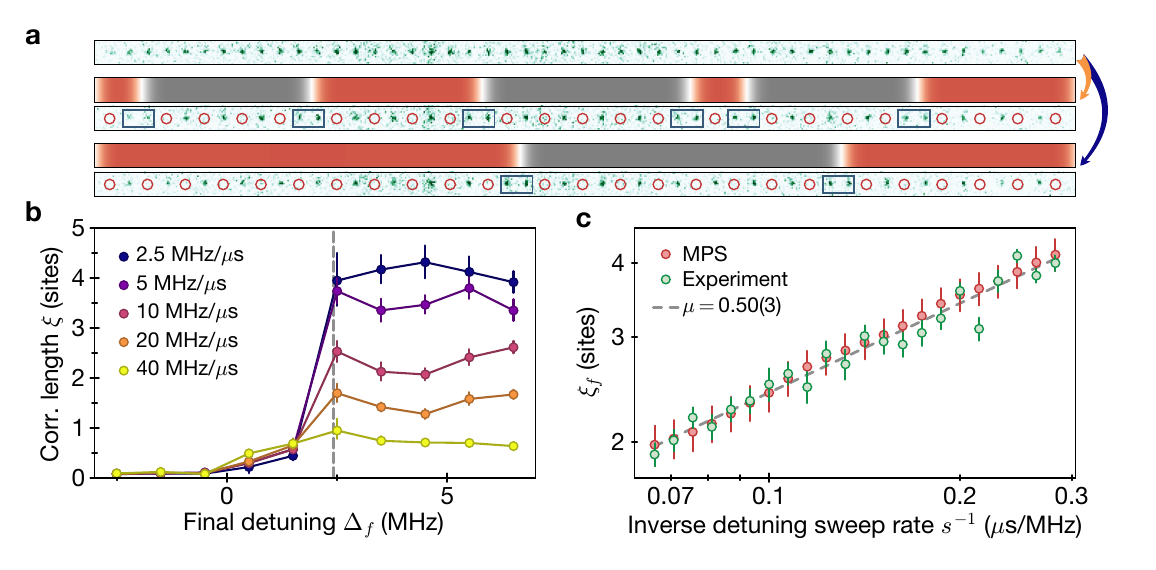}
	\caption{
	  \textbf{Quantum Kibble-Zurek mechanism for a quantum phase transition (QPT) into the $\mathbb{Z}_2$-ordered phase.} 
	\textbf{a}, Single-shot images of the atom array before and after a fast (orange arrow) and a slow (blue arrow) sweep across the phase transition, showing larger average sizes of correlated domains for the slower sweep. Green spots (open circles) represent atoms in $\ket{g}$ ($\ket{r}$). Blue rectangles mark the position of domain walls, and the red and gray colored regions above highlight the extent of the correlated domains. \textbf{b}, Correlation length growth and saturation as the system crosses the QPT at different rates. The gray dashed line indicates the critical detuning. \textbf{c}, Dependence of correlation length on inverse sweep rate across the phase transition with experimentally measured (green) and matrix product state-simulated results (red). The length is extracted from fitting the modulus of the correlation data to an exponential decay. Error bars denote fit uncertainty. The dashed line indicates a power-law fit with a scaling exponent $\mu = 0.50(3)$ for the experiment.}
	\label{fig:Correlations}
\end{figure*}

The celebrated Kibble-Zurek mechanism~\cite{Kibble1976,Zurek1985} describes nonequilibrium dynamics and the formation of topological defects in a second-order phase transition driven by thermal fluctuations, and has been experimentally verified in a wide variety of physical systems~\cite{delCampo2014,
Navon2015}. Recently, the concepts underlying the Kibble-Zurek description have been extended to the quantum regime~\cite{Polkovnikov2005, Zurek2005, Dziarmaga2005}. Here, the typical size of the correlated regions, $\xi$, after a dynamical sweep across the QPT scales as a power-law of the sweep rate, $s$, with an exponent, $\mu$, determined entirely by the QPT's universality class. Specifically, QKZM  postulates that when the time scale over which the Hamiltonian changes becomes faster than the characteristic response time,  $\tau$, determined by the inverse of the energy gap between the ground and excited states, nonadiabatic excitations prevent the continued growth of  correlated regions (Figs.~\ref{fig:Phases}a,b). The resulting scaling exponent, $\mu = \nu /(1+\nu z)$, is determined by a combination of the critical exponent $\nu$, that characterizes the divergent correlation length, and the dynamical critical exponent $z$, that characterizes the relative scaling of space and time close to the critical point \cite{Sachdev2009}.  While QKZM 
has many important implications, e.g. in 
quantum information science~\cite{Farhi2000}, 
its experimental verification 
is challenging 
due to the coupling of many-body systems to the environment
~\cite{Gardas2018}. Recently, experimental control over isolated quantum systems enabled the observation of scaling behavior across quantum phase transitions described by mean-field theories~\cite{Anquez2016,Clark2016}. While important aspects of QPTs have already been explored in strongly correlated systems~\cite{Endres2012}, experimental observation of quantum critical phenomena beyond mean-field in real-time dynamics remains an outstanding challenge~\cite{Chen2011,Braun2015,Gardas2018}.

We probe quantum criticality using a reconfigurable 1D array of $\Rb$ atoms with programmable interactions~\cite{Bernien2017}. In our system, 51 atoms in the electronic ground state $\ket{g}$, evenly separated by a controllable distance, are homogeneously coupled to the excited Rydberg state $\ket{r}$, in which they experience van der Waals interactions with a strength that decays as $V(r)\propto1/r^6$, where $r$ is the interatomic distance. This system is described by the many-body Hamiltonian,
\begin{equation}
	\label{eq:Rydberg-Hamiltonian}
	\frac{\ham}{\hbar} = \frac{\Omega}{2}\sum_i(\ket{g_i}\bra{r_i} + \ket{r_i}\bra{g_i})-\Delta\sum_i  n_i +\sum_{i<j}V_{ij}n_in_j,
\end{equation}
where $n_i = \ket{r_i}\bra{r_i}$ is the projector onto the Rydberg state at site $i$, $\Delta$ and $\Omega$ are the detuning and Rabi frequency of the coherent laser coupling between $\ket{g}$ and $\ket{r}$, and $V_{ij}$ is the interaction strength between atoms in the Rydberg state at sites $i$ and $j$. For negative values of $\Delta$, the many-body ground state corresponds to a state in which all atoms are in the electronic ground state $\ket{g}$, up to quantum fluctuations, and belongs to a so-called ``disordered'' phase with no broken spatial symmetry. For $\Delta>0$, several spatially ordered phases arise from the competition between the detuning term, which favors a large Rydberg fraction, and the Rydberg blockade, which prohibits simultaneous excitation of atoms separated by a distance smaller than the blockade radius, $R_B$, defined via $V(R_B) \equiv \Omega$.  As illustrated in Fig.~\ref{fig:Phases}c,d, we probe different QPTs into states breaking various symmetries by choosing the interatomic spacing, and sweeping the control parameter, $\Delta$, across the phase boundary. 

\begin{figure*}
	\includegraphics{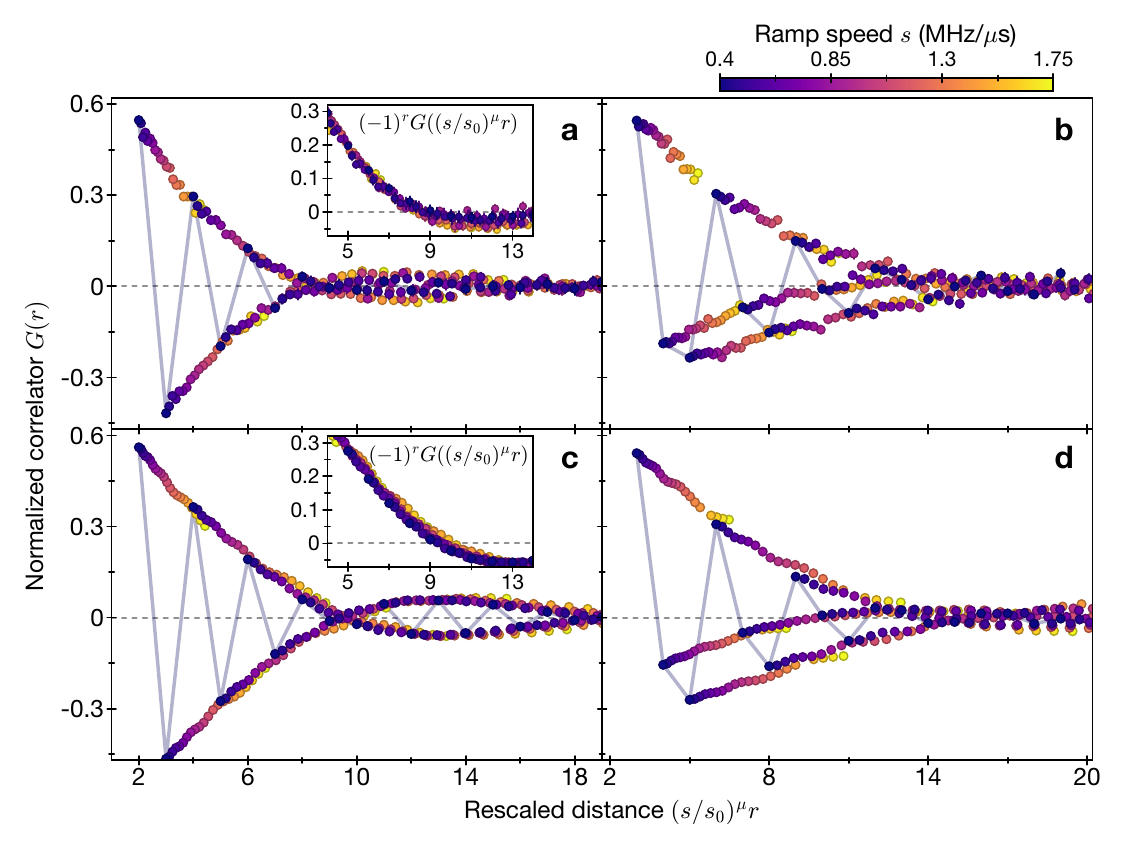}
	\caption{\textbf{Universality of spatial correlations.}
    Collapse of the measured  (\textbf{a}) and numerically calculated 
(\textbf{c}) correlations in the $\mathbb{Z}_2$-ordered phase with distances rescaled according to the extracted scaling exponents. The blue line connects the points of the correlation function corresponding to the slowest sweep rate. The insets show the staggered rescaled correlations. The negative values of the correlation function indicate nontrivial correlations between domain walls. Collapse of the measured (\textbf{b}) and numerically calculated (\textbf{d}) correlations in the $\mathbb{Z}_3$-ordered phase highlighting the energetic difference of the different types of defects, as shown by the distinguishability of the two negative branches, i.e., a deviation from a period-3 density wave. All error bars indicate the s.e.m.}
	\label{fig:Universality}
\end{figure*}
 
\begin{figure*}
	\includegraphics{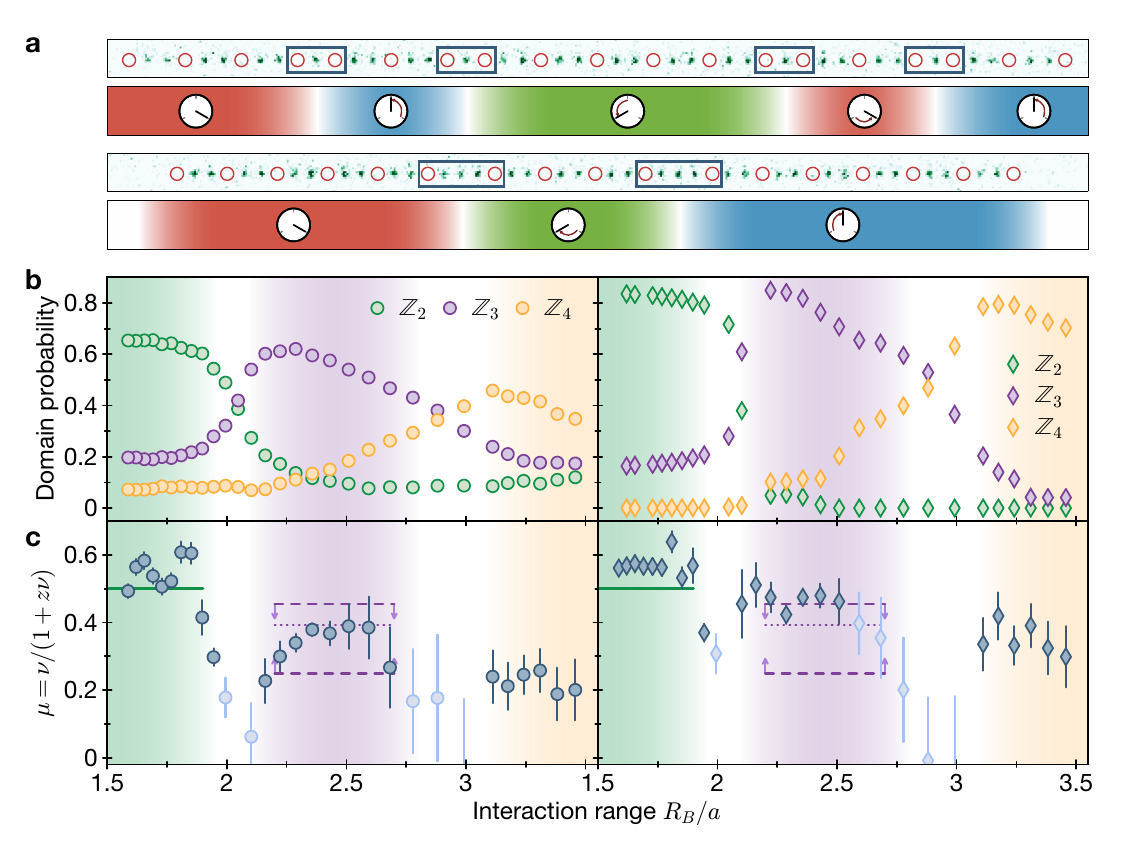}
	\caption{
        \textbf{Power-law scaling for different interactions.
        }
		\textbf{a}, Experimental realization of the chiral clock model~\cite{Samajdar2018}. The top row shows a single fluorescence image of a state in the $\mathbb{Z}_3$-symmetry broken phase ($R_B/a\sim2.16$), with four $\mathbb{Z}_2$-type defects displacing the Rydberg atoms in one direction (counter-clockwise chirality). The bottom rows display a system with stronger interactions ($R_B/a\sim2.43$), where $\mathbb{Z}_4$-type defects are favored, and the Rydberg atoms are displaced in the opposite direction (clockwise chirality). The colored regions highlight the extent of the correlated domains, labeled by clock orientations in connection to the chiral clock model. \textbf{b}, Fraction of the final state consistent with the different $\mathbb{Z}_N$-ordered states observed in the experiment (left, circles) and in numerical simulations (right, diamonds). Within the $\mathbb{Z}_3$-ordered region, the most dominant type of defect changes from $\mathbb{Z}_2$- to $\mathbb{Z}_4$-type as the interaction range increases. The higher contrast in the calculated domain probabilities in Fig.~\ref{fig:Scaling}b  is  due to finite detection fidelity, which does not affect the extracted value of $\mu$. \textbf{c}, Scaling exponent, $\mu$, as a function of $R_B/a$ obtained from experimental data (left, circles), and matrix product state simulations (right, diamonds). Pale blue points indicate instances where the measured correlation lengths do not grow beyond the size of $R_B/a$.
        Shaded areas indicate the regions consistent with $\mathbb{Z}_2$- (green), $\mathbb{Z}_3$- (purple), and $\mathbb{Z}_4$-ordered (orange) phases. The solid green line corresponds to $\mu_{\textsc{Ising}}$, the purple dashed lines represent the upper~\cite{Samajdar2018}, and lower~\cite{Whitsitt2018} bounds of $\mu_{\textsc{CCM}}$, while the purple dotted line is the value of $\mu_{\textsc{CCM}}$ obtained from the best numerical estimates of $z$~\cite{Samajdar2018} and $\nu$~\cite{Chepiga2018}. Error bars represent the 68\% confidence interval (b), and uncertainty of the power-law fit (c), which is dominated by systematic effects in the extraction of individual correlation lengths.
}
	\label{fig:Scaling}
\end{figure*}

We first focus on the QPT 
 into the antiferromagnetic phase with broken $\mathbb{Z}_2$ symmetry, which is known to belong to the Ising universality class~\cite{Sachdev2009}. Using an interatomic spacing, $a$, such that  $R_B/a \sim 1.69$, we create an array of 51 atoms in the electronic ground state, and 
slowly turn on $\Omega$ at $\Delta<0$, adiabatically preparing the system in the ground state of the disordered phase. The detuning is then increased at a constant rate, $s$, up to a final value $\Delta_f$, at which point $\Omega$ is slowly turned off 
(see inset of Fig.~\ref{fig:Phases}c), and the state of every atom is measured. We examine the dynamical development of correlations between the atoms, characterized by the Rydberg density-density correlation function:
\begin{equation}
\label{eq:Correlation_function}
	G(r) = \sum_i(\langle n_{i}n_{i+r}\rangle - \langle n_i\rangle\langle n_{i+r}\rangle)/N_r,
\end{equation}
where the normalization $N_r$ is the number of pairs of sites separated by distance $r$. By fitting an exponential decay to the modulus of the correlation function, we extract the correlation length. The experimental results show growth of the correlation length as the detuning approaches the critical point, followed by saturation once the detuning is swept past the critical point into the ordered phase (Fig. 2b). From the individual  images, it is apparent that, while for fast sweeps the ordered domains are frequently interrupted by defects (domain walls), for slow ramps, significantly longer domains are observed (Fig.~\ref{fig:Correlations}a).
A systematic analysis of the final correlation lengths after crossing into the ordered phase shows that a power-law scaling model $\xi(s) = \xi_{0}(s_{0}/s)^{\mu}$ with $\mu = 0.50(3)$ accurately describes our measurements (Fig.~\ref{fig:Correlations}c).  These results are consistent with numerical simulations (red points) of the coherent evolution of the system using Matrix Product States (MPS).

The QPT into the $\mathbb{Z}_2$-ordered phase is in the Ising universality class~\cite{Sachdev2009}, with critical exponents in 1D of $z =1$, $\nu = 1$, and consequently, $\mu_{\textsc{Ising}} = 0.5$. Our observations are 
consistent with these quantitative predictions, and are quite distinct from those associated with a mean-field Ising transition, described by $z = 1$, $\nu = \nicefrac{1}{2}$, and yielding $\mu_{\textsc{mf}} = \nicefrac{1}{3}$~\cite{Sachdev2009,Anquez2016}. These results offer the first experimental verification of the quantum Kibble-Zurek mechanism in an isolated quantum system that defies a mean-field description. 

A key concept associated with critical phenomena is that of universality, which is manifested by the collapse of correlations to a universal form when rescaled according to the corresponding critical exponents~\cite{Sachdev2009}. Such a signature is a strong test of an underlying universal scaling law, and in connection with the QKZM, should appear upon rescaling lengths by $(s/s_0)^{\mu}$~\cite{Kolodrubetz2012}.  Fig.~\ref{fig:Universality}a shows that the rescaled correlations for $R_B/a \sim 1.81$ indeed collapse onto two smooth branches, which in turn collapse on top of each other when the correlations are rectified as $(-1)^{r}G(r)$ (inset in Fig.~\ref{fig:Universality}a), according to the $\mathbb{Z}_2$ order parameter.

While the quantum Kibble-Zurek mechanism is a coarse-grained description predicting the mean density of defects, the shape of the correlation function gives further access to microscopic details of the system. Detailed inspection of the rescaled correlation functions reveals nontrivial deviations from a simple exponential decay. In particular, the correlations in Fig.~\ref{fig:Universality}a become negative for a range of distances, which implies complex dynamics in the formation and spreading of defects. 
The observed corrections to simple QKZM predictions are consistent with recent theoretical analyses~\cite{Kolodrubetz2012,Cherng2006} and are in good agreement with numerical simulations using MPS (Fig.~\ref{fig:Universality}c). 
Finally, applying the universal rescaling to the correlation growth shown in Fig.~\ref{fig:Correlations}b allows us to independently estimate the values of critical exponents (Extended Data Fig. 7), showing that our results are consistent with $z=\nu=1$ associated with the Ising QPT. 

Having established the validity of the QKZM, as well as its limitations, for a QPT in the  Ising universality class, we now explore transitions into more complex  $\mathbb{Z}_N$-ordered phases, where Rydberg excitations are evenly separated by $N>2$ sites (see Fig.~\ref{fig:Phases}c).
The correlation functions at smaller interatomic spacings after slow detuning sweeps 
reflect the spatial ordering of the $\mathbb{Z}_3$- and $\mathbb{Z}_4$-ordered phases (Fig.~\ref{fig:Phases}d).
In addition, we determine the probability of finding two Rydberg excitations separated by $N$-sites, for each value of $N$ and $R_B$ (Fig.~\ref{fig:Scaling}b). Combining these measurements with the numerically obtained critical points (see Fig.~\ref{fig:Phases}c), we experimentally identify approximate boundaries for the regions consistent with the $\mathbb{Z}_2$-, $\mathbb{Z}_3$-, and $\mathbb{Z}_4$-ordered phases in Fig.~\ref{fig:Scaling}b. Within these regions, the dominant type of order is the one associated with the corresponding phase, while the second most prevalent type of order arises from the lowest-energy (most probable) defects. In particular, we observe that in the $\mathbb{Z}_3$-ordered phase, the most-likely defect changes from $\mathbb{Z}_2$-like for smaller values of $R_B/a$, to $\mathbb{Z}_4$-like as $R_B/a$ increases. 

We test for a power-law scaling behavior of the correlation length growth as a function of ramp speed at different interaction strengths in  Fig.~\ref{fig:Scaling}c. To consistently compare the results for all interaction strengths, we fit the correlation function to an exponentially decaying density wave with a period set by the underlying order (as opposed to the modulus of the correlation function used in Fig.~2c). 
The scaling is extracted through a power-law fit to the resulting correlation lengths. In parameter regimes far away from regions of competing order, we observe three stable plateaus for the regions consistent with $\mathbb{Z}_2$, $\mathbb{Z}_3$, and $\mathbb{Z}_4$ order, respectively. For interaction strengths where there is a strong competition between different types of order, we do not observe the formation of long-range correlations (pale points in Fig.~\ref{fig:Scaling}c). In these cases, the detuning sweeps either do not fully cross the phase boundary into the ordered phases (Methods) or potentially enter theoretically predicted incommensurate phases~\cite{Ostlund1981,Fendley2004}.

To understand these observations, we compare them to finite-size scaling analyses of ground-state properties~\cite{Samajdar2018, Whitsitt2018, Chepiga2018}, as well as MPS-based numerical simulations of our experimental protocol
for the full Hamiltonian~\eqref{eq:Rydberg-Hamiltonian}. For the transitions into the $\mathbb{Z}_2$-ordered phase, some of the extracted values of $\mu$ are slightly larger than the expected exponent from the Ising model 
$\mu_{\text{Ising}}=0.5$. We attribute these deviations, 
to a combination of the long-ranged interactions, finite-size and/or time effects, and systematic effects related to the inversion of the alternating pattern (Fig.~\ref{fig:Universality}a,c, see also Methods). 

Quantum phase transitions associated with the breaking of a $\mathbb{Z}_3$ symmetry are more complex due to competition between the different types of 
defects that can be formed. In our system, the defects correspond to two different types of domain walls, where the distance between neighboring Rydberg excitations is $2$ sites and $4$ sites (see Fig.~4a). For the experimentally accessible parameter regimes, the different associated excitation energies lead, in general, to an asymmetry between these defects (see also Fig.~\ref{fig:Scaling}b). Correspondingly, the $\mathbb{Z}_3$-symmetry breaking is believed to be in the universality class of the $3$-state \textit{chiral} clock model (CCM) (Fig.~\ref{fig:Scaling}a, Methods, and~\cite{Samajdar2018}). 

The exact nature of such phase transitions has been a subject of intense theoretical research for the past three decades~\cite{Huse1982,Ostlund1981,haldane1983phase,Samajdar2018,Whitsitt2018,Chepiga2018}. Only very recently, numerical studies of equilibrium scaling properties~\cite{Samajdar2018,Whitsitt2018,Chepiga2018} 
provided evidence for 
a direct transition~\cite{Chepiga2018} along some paths across the phase boundary, where  
the expected range of values of the scaling exponent is $\mu<0.45$~\cite{Samajdar2018}, and $\mu>0.25$~\cite{Whitsitt2018}.
Our experimental results are 
consistent with a direct CCM phase transition over a range of interaction strengths
with $\mu\sim0.38$,  in agreement with the theoretical value obtained by combining the results of the most extensive numerical finite-size scaling studies \cite{Samajdar2018,Chepiga2018} (dashed line in Fig. 4c). Further evidence for a direct chiral QPT is provided by 
the universal scaling behavior  into the $\mathbb{Z}_3$-ordered phase (see Fig.~\ref{fig:Universality}b,d). 

The transition into the $\mathbb{Z}_4$-ordered phase is even more involved. At present, complete understanding of this transition is lacking, in particular  
due to the potential presence of an
intermediate gapless incommensurate phase~\cite{Ostlund1981,haldane1983phase}.  Our experimental results in this region are reasonably consistent with power-law scaling with $\mu~\sim~0.25$.  While recent theoretical work shows that QKZM scaling may still hold on quenching through a gapless phase, albeit with a modified (system-specific) power-law  exponent~\cite{Dutta2015}, 
detailed theoretical understanding of our experimentally observed exponents in the $\mathbb{Z}_4$ regime requires further studies.

Detailed comparison of  our experimental results across all phases to the numerical simulation of the Hamiltonian dynamics using MPS are presented in Figure 4. While 
 qualitatively similar, they display clear discrepancies. 
 Most significant is  
 a systematic offset in the extracted values of $\mu$
 between experiment,  finite-size scaling analysis and time-dependent MPS simulations. 
 While it  can be potentially attributed to experimental imperfections and 
 subtle differences between the experimental system and the model used for the numerical simulations (see Methods), the disagre.ement of MPS with both experimental results and finite-size scaling analyses of equilibrium properties
highlights the difficulty in approximately modeling complex nonequilibrium dynamics of many-body systems. 

Our observations demonstrate a novel approach for probing quantum critical phenomena and provide new insights into the physics of exotic QPTs that do not lend themselves to simple theoretical analyses. 
Increasing the system size, improving atomic coherence properties, and exploring wider parameter regimes may allow for more precise probing of exotic QPTs into both ordered and incommensurate phases \cite{Ostlund1981,Fendley2004,Samajdar2018,Chepiga2018} in various models. In particular, the present approach is well suited for simulations of lattice gauge theories~\cite{Weimer2010_b}.  Whereas the system studied here is formally equivalent 
to a quantum link model on a ladder~\cite{Moessner2001},   two- and three-dimensional  systems, realized  using novel trapping techniques~\cite{Barredo2018, Kumar2018}, can be used to simulate a wide variety of non-trivial lattice gauge models~\cite{Tagliacozzo2013}.  
Finally, the methods demonstrated  in this work can be used to effectively encode and explore solutions to computationally difficult combinatorial optimization  problems such as finding the Maximum Independent Set~\cite{Pichler2018}. Detailed understanding of quantum dynamics in such systems might have direct applications for exploring quantum speedup in both adiabatic and dynamical quantum optimization algorithms~\cite{Farhi2000}.  

\FloatBarrier
\renewcommand{\addcontentsline}[3]{}

\bibliographystyle{naturemag}

\smallskip\noindent\textbf{Acknowlegements}
We thank Anushya Chandran, Eugene Demler,  Anatoli Polkovnikov, and Ashvin Vishwanath for insightful discussions. This work was supported by NSF, CUA, ARO,  AFOSR MURI, DOE, and Vannevar Bush Faculty Fellowship. A.O. acknowledges support by a research fellowship from the German Research Foundation (DFG). H.L. acknowledges support from the National Defense Science and Engineering Graduate (NDSEG) fellowship. S. Schwartz acknowledges funding from the European Union under the Marie Sk\l odowska Curie Individual Fellowship Programme H2020-MSCA-IF-2014 (project number 658253). H.P. acknowledges support by the National Science Foundation (NSF) through a grant at the Institute of Theoretical Atomic Molecular and Optical Physics (ITAMP) at Harvard University and the Smithsonian Astrophysical Observatory. ME acknowledges funding provided by the Institute for Quantum Information and Matter, an NSF Physics Frontiers Center (NSF Grant PHY-1733907). S. Sachdev acknowledges support from the US Department of Energy (grant number DE-SC0019030).

\smallskip\noindent\textbf{Author Contributions}
The experimental measurements and data analysis were carried out by A.K., A.O.,H.L., and H.B..
Theoretical analysis was performed by H.P., S.C., and R.S..
S.Schwartz, P.S., S.Sachdev, P.Z., and M.E. contributed to the development of measurement protocols and theoretical models, and the interpretation of results.
All work was supervised by M.G., V.V. and M.D.L..
All authors discussed the results and contributed to the manuscript.

\smallskip\noindent\textbf{Correspondance and requests for materials} 
should be addressed to M.D.L.


\clearpage
\newpage

\setcounter{figure}{0}
\setcounter{table}{0}
\makeatletter 
\renewcommand{\thefigure}{Extended Data Fig. \@arabic\c@figure}
\renewcommand{\fnum@figure}{\textbf{Extended Data Figure \@arabic\c@figure}}
\renewcommand{\thetable}{Extended Data Table \@arabic\c@table}
\renewcommand{\fnum@table}{\textbf{Extended Data Table \@arabic\c@table}}
\makeatother

\section{Methods}

\noindent\textbf{Rydberg array preparation.} The experiment utilizes an acousto-optic deflector to generate multiple optical tweezers, which are loaded probabilistically from a cold gas of \Rb atoms in a magneto-optical trap. Each tweezer can be loaded with up to a single atom. Once the cloud is dispersed, a fluorescence image, similar to the ones shown in Fig.~2a of the main text, is taken to identify loaded traps. The traps are then rearranged to generate a defect-free regular array of 51 atoms, evenly separated by a distance $a$~\cite{Endres2016}.

We define our spin Hamiltonian according to two pseudospin-$1/2$ states. The first is a ground-state hyperfine sublevel, $\ket{g} = \ket{5S_{1/2}, F=2, m_F=-2}$. The second is the interacting Rydberg state ${\ket{r}=\ket{70S, J=1/2, m_J=-1/2}}$. These two states are coupled by a two-photon process via the intermediate state $\ket{e} = \ket{6P_{3/2}, F=3, m_F=-3}$. The two lasers operate at wavelengths $420\,$nm for the lower transition and $1013\,$nm for the upper transition.

The $420\,$nm laser is a frequency-doubled Titanium-Sapphire laser (SolsTiS 4000 PSX F by M Squared), locked to an optical reference cavity (ATF-
6010-4 from Stable Laser Systems). The $1013\,$nm laser is an external cavity diode laser (CEL002 by MOGLabs) that is locked to the same reference cavity. The transmitted light through the cavity is used to injection lock another $1013\,$nm laser diode, which is then amplified by a tapered amplifier~\cite{Levine2018}. 

Both beams are focused along the array axis (aligned with the quantization axis) to drive $\sigma^-$ and $\sigma^+$ transitions for the $420\,$nm and $1013\,$nm beams, respectively.\pargap
\noindent\textbf{Pulse generation.} We modulate the $420\,$nm Rydberg laser with an AOM driven by an arbitrary waveform generator (AWG, M4i.6631-x8 by Spectrum Instrumentation). For each experiment, we program a waveform with varying amplitudes, frequencies and phases in the time domain into the AWG, which is then transmitted to the AOM through a high-power RF amplifier (ZHL-1-2W+ by Mini-Circuits).

\begin{figure*}[t]
  \centering
  \includegraphics[width=0.5\linewidth]{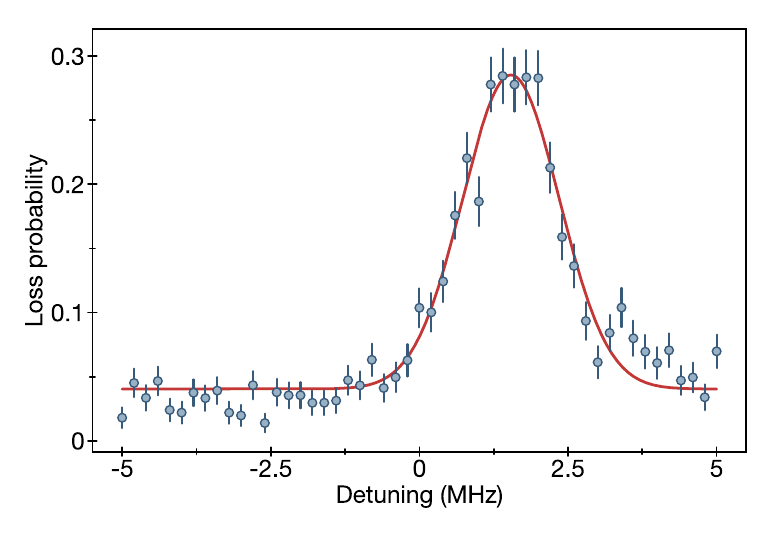}
  \caption{\textbf{Determination of initial detuning $\Delta_{0}$.} At a fixed laser detuning, we linearly ramp $\Omega$ on and then off over $1~\mu$s each. We identify the negative detuning closest to resonance for which we are fully adiabatic, such that the excitation probability at the end of the pulse returns to the minimum. From this typical measurement, taken at $R_B/a = 1.59$, we set $\Delta_0 = -2.5\,$MHz. Error bars denote 68\% confidence intervals.}
  \label{sfig:Delta0}
\end{figure*}

The nonlinear AOM response to changes in amplitude and frequency poses a technical challenge. The deflection efficiency is not proportional to the waveform amplitude, and large changes in the waveform frequency lead to variations in the deflection efficiency. These effects lead to distortions in the pulse shape.  
We apply feed-forward corrections to the amplitude to both match the output intensity to the desired waveform amplitude, as well as to compensate for the variations with frequency.
\pargap
\noindent\textbf{Pulse Parameters.} All pulses begin by turning on the value of $\Omega$ linearly over $1\,\mu$s at a fixed initial detuning $\Delta_{0}$. We select our initial detuning to be as close to the critical point as possible subject to the constraint that the initial turn-on is still fully adiabatic. We identify this detuning experimentally by ramping $\Omega$ on and then off for various fixed detunings. In the adiabatic case, all the atoms should return to $\ket{g}$. We therefore select the detuning closest to resonance that still shows no excess excitation at the end of the pulse. For a typical measurement in the $\mathbb{Z}_2$ regime, we select $\Delta_0 = -2.5\,$MHz (\ref{sfig:Delta0}).

\begin{figure*}[t]
  \centering
  \includegraphics[width=0.7\linewidth]{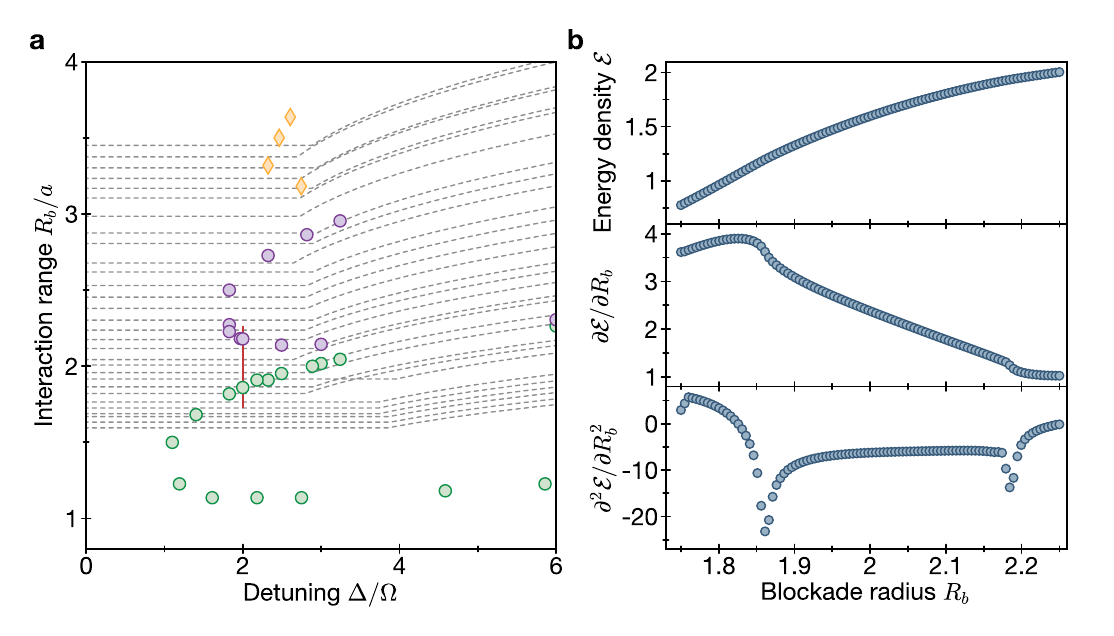}
  \caption{\textbf{Numerically extracted phase diagram with trajectories for QKZM measurements.} \textbf{a}, Green (purple) markers indicate the phase boundary points between disordered and $\mathbb{Z}_2$-($\mathbb{Z}_3$-)ordered phases. Yellow diamonds indicate the boundaries of the disordered phase (as approached from increasing $\Delta$ with fixed $\Omega$ and $R_b/a$). We have not verified if these transitions are directly from disordered to $\mathbb{Z}_4$-ordered phases, or involve incommensurate phases. Each gray dashed line corresponds to the trajectory across phase space used to probe for scaling behavior of correlation length growth. The horizontal section of each trace corresponds to the detuning sweep at a constant Rabi frequency, while the curved sections correspond to pulse turn-off at a fixed value of the detuning. The total duration of the detuning sweep is varied to control the rate of transition across the phase boundaries, but the time to turn the field off is not. \textbf{b}, Numerically obtained energy densities $\mathcal{E}$ along the red solid line indicated in \textbf{(a)}. The second order derivatives of $\mathcal{E}$ shows clear cusps at two critical points.}
  \label{sfig:PhaseDiagramTrajectories}
\end{figure*}

The final detunings of the sweeps are chosen in most cases to cross the tip of the corresponding phase boundary. In some cases in which the interaction strength is on the border between two phases, we do not fully cross over the boundary (\ref{sfig:PhaseDiagramTrajectories}a).

The power-law scaling behavior of the correlation length can be limited owing to strong nonadiabaticity far away from the critical point, where the behavior of the system is susceptible to the microscopic details and should deviate from universal theories, limiting how fast a sweep across the phase transition can be. At the same time, slow sweeps are more susceptible to decoherence, both because of the longer pulse time window, and because the system remains closer to the ground state near the critical point and the growing quantum correlations are increasingly sensitive to environmental noise. To determine the range of rates for which QKZM scaling can be observed, we perform a sweep into each of the ordered phases at a wide range of sweep speeds $s$. We fit the correlation lengths for each parameter, discarding all the instances where the correlation length is smaller than the size of the blockade radius, to a model that accounts for incoherent processes as a saturation in the final size of the correlation length, namely:
\begin{equation}
\xi(s) = \begin{cases}
\xi_{0}(s_{0}/s)^{\mu} &: s\leq s_{c},\\
\xi_{0}(s_{0}/s_{c})^{\mu} &: s>s_{c}.
\end{cases}
\end{equation}

\begin{figure*}[t]
  \centering
  \includegraphics[width=0.5\linewidth]{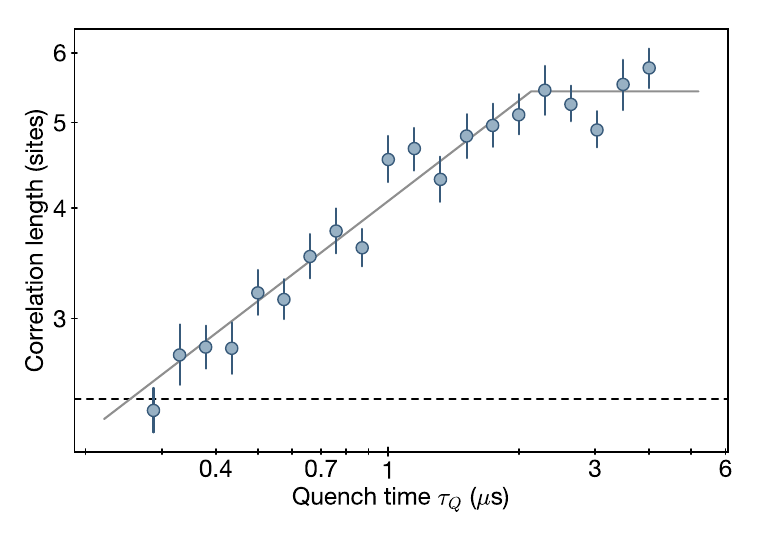}
  \caption{\textbf{Scaling window.} Determination of the window of rates where scaling is valid for the transition into the $\mathbb{Z}_3$-ordered phase. The black solid lines represent the result of the fitted model which grows as a power law until it saturates. The dashed horizontal line marks the size of the blockade radius. The values of all the rates used in the experiment are larger than the value at which the dashed and solid lines intersect, and smaller than the point where the model saturates. The error bars denote the uncertainty of the power-law fit.}
  \label{sfig:rate_window}
\end{figure*}

From this fit, we set $s_{\rm min} > s_{c}$ and find $s_{\rm max}$ such that $\xi(s_{\rm max}) > R_{B}$. An example of this can be seen in \ref{sfig:rate_window}. In this way, we determine the sweep parameters for the different values of the interaction strength (see \ref{table:exp_parameters}).
\pargap
\noindent\textbf{Numerical computation of the phase diagram.} The quantum critical points along the phase boundary on the phase diagram presented in the main text were obtained using both finite- and infinite-system density-matrix renormalization group (DMRG) algorithms~\cite{white1992density, white1993density,ostlund1995thermodynamic, rommer1997class, dukelsky1998equivalence,peschel1999density}. The filled colored regions are not the result of numerical simulations, and only show approximately the expected shape of the phases. In this section, we describe the details of the DMRG calculations.

\begin{figure*}[htb]
  \centering
  \includegraphics[width=0.4\textwidth]{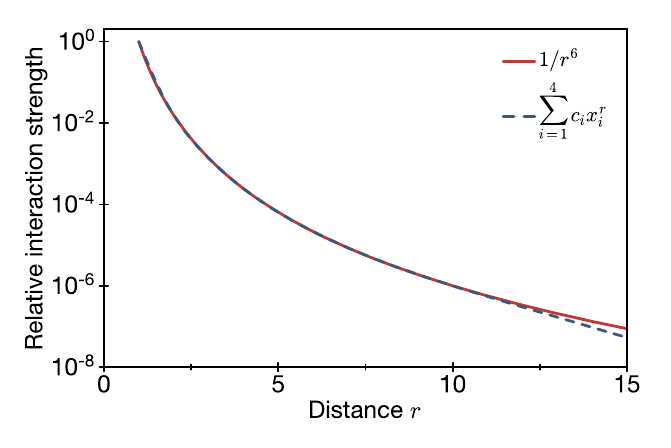}
  \caption{\textbf{Interaction potential approximation.} Comparison between the exact power-law decay $1/r^6$ and its approximation using a linear combination of four exponentials. The two functions agree with each other until their relative strength decreases to $10^{-6}$.}
  \label{sfig:power_law_apprx}
\end{figure*}

For the infinite-system DMRG (iDMRG), we generally follow the method summarized in Ref.~\cite{McCulloch2008}, where translationally invariant matrix product states (iMPS) are used as variational ansatze for ground-state wavefunctions. Our Hamiltonian with long-range interactions is encoded using matrix product operator representations, where $1/r^6$ decaying interactions are approximated by a linear combination of four exponentials
\begin{align}
\frac{1}{r^6} \approx \sum_{i=1}^4 c_i x_i^r,
\end{align}
with $(c_1, c_2, c_3, c_4) = (170.55, 1.29, 0.0252, 0.000279) $ and $(x_1, x_2, x_3, x_4) = (0.00519, 0.0835, 0.279, 0.565) $~\cite{Pirvu2010}. The resultant function provides an excellent approximation with relative error less than $10^{-5}$ (\ref{sfig:power_law_apprx}). This accuracy implies that even with the strongest interaction strength probed in our experiments ($R_b \approx 3.5$), the maximum correction, $V_0 \left|1/r^6 - \sum_{i=1}^4 c_i x_i^r\right| \lesssim (2\pi)\times 36$~kHz, is much weaker than the smallest energy scale that can be probed within our experimental timescales.

Our phase diagram involves quantum phases that spontaneously break spatial translation symmetry. Hence, it is important that the number of spins in a translationally invariant unit cell of our iMPS ansatz must be compatible with the broken spatial symmetry. We use 2 or 6 spins as a unit cell in order to probe phase transitions from disordered to $\mathbb{Z}_2$-ordered or $\mathbb{Z}_3$-ordered phases, respectively.
Incommensurate phases or onset of spatial symmetry breaking that is not compatible with the number of spins per unit cell can be identified by oscillatory behavior of wavefunction overlaps or energy densities over iterations.

In order to obtain the ground-state wavefunction, we iteratively optimize iMPS tensors until the (local) overlap between wavefunctions from two consecutive optimization steps approaches unity up to a small error $\epsilon$. As convergence criteria, we require that either $\epsilon \leq 10^{-8}$ or $\epsilon$ is limited by truncation errors arising from finite bond dimension $D$~\cite{McCulloch2008}. We use a wide range of bond dimensions up to $D = 200$, depending on the quantity of interest to be computed and on the convergence of wavefunctions. For example, computing the ground state energy density is relatively insensitive to bond dimensions, while extracting correlation lengths near the critical point requires  a substantially larger $D$.

We thus extract the phase boundaries from the energy density. Specifically, we use iDMRG to extract the ground-state energy density $\mathcal{E}$ along a line in the parameter space, $(R_b/a, \Delta/\Omega)$, and compute its second derivative along the line. When crossing a quantum phase transition, the second-order derivative of the energy density exhibits a sharp feature. For example, \ref{sfig:PhaseDiagramTrajectories}b shows the numerically computed energy densities per unit cell (6 spins) as a function of $R_b/a \in [1.75, 2.25]$ for a fixed $\Delta /\Omega = 2$ with $D = 10$. We find clear cusps at $R_b/a \approx 1.86$ and $2.18$, corresponding to critical points from $\mathbb{Z}_2$-ordered to disordered and to $\mathbb{Z}_3$-ordered phases. Similar procedures along different lines lead to the phase diagram in \ref{sfig:PhaseDiagramTrajectories}a and in Fig. 1c of the main text.

\begin{figure*}[htb]
  \centering
  \includegraphics[width=0.5\linewidth]{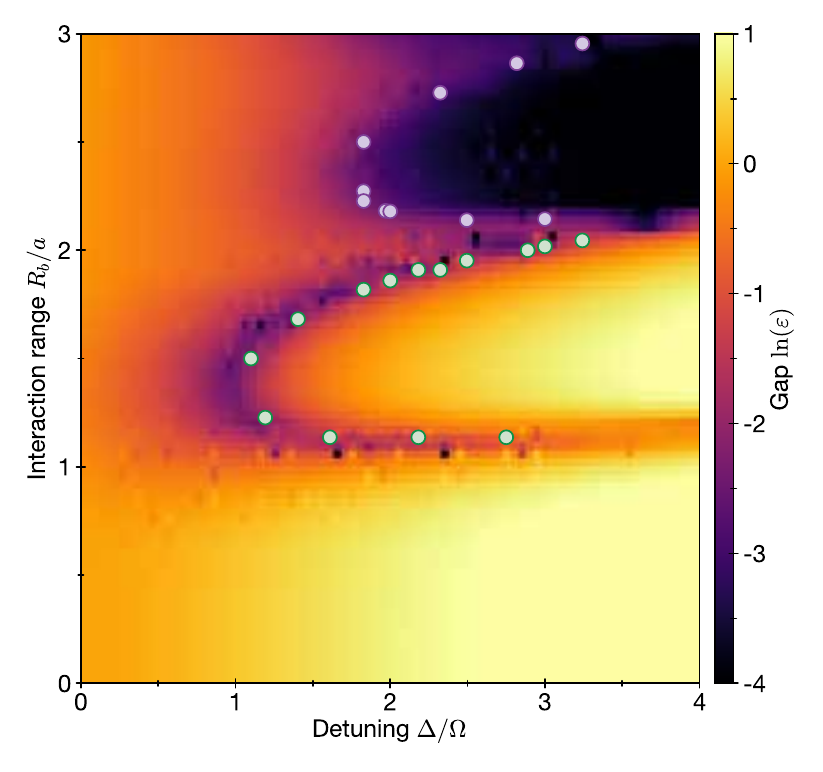}
  \caption{\textbf{Energy gap.} Calculated gap between ground and first excited state using density-matrix renormalization group (DMRG) calculations. Green (purple) circles indicate the extracted quantum critical points separating the disordered from the $\mathbb{Z}_2$($\mathbb{Z}_3$)-ordered phase.}
  \label{sfig:energy_gap}
\end{figure*}

These phase boundaries are also reproduced using finite-system DMRG~\cite{schollwock2005density, schollwock2011density} with a bond dimension up to $D = 60$ for a chain of $L=51$ atoms and open boundary conditions. The first three energy levels are individually targeted, which, in turn, gives us access to the energy gap. The closing of the gap outlines well-defined lobes in the phase diagram, the boundaries of which overlap well with the points extracted previously with iDMRG (see \ref{sfig:energy_gap}).

A few remarks are in order. First, it has been previously discussed that the $\mathbb{Z}_3$-ordered phase may be interfacing incommensurate phases~\cite{Fendley2004}. However, we do not find any evidence of incommensurate phases between $\mathbb{Z}_2$ and $\mathbb{Z}_3$ phases up to $\Delta/\Omega = 12$ within our numerical precision. The nature of the direct transition from disordered to $\mathbb{Z}_3$-ordered phases is discussed in Refs.~\cite{Samajdar2018, Whitsitt2018, Chepiga2018}. Second, we have not explicitly identified the phase transition between disordered to $\mathbb{Z}_4$-ordered phases. This is because our choices of a unit cell (two or six spins) are not compatible with $\mathbb{Z}_4$-ordered wavefunctions. Instead, the boundary of the disordered phase for $R_b/a > 3$ (yellow diamonds in \ref{sfig:PhaseDiagramTrajectories}a) has been extracted from the convergence of the iDMRG algorithm; as $\Delta/\Omega$ increases with a fixed $R_b/a$, the yellow diamonds in \ref{sfig:PhaseDiagramTrajectories}a indicate the points at which the iDMRG algorithm ceases to converge, and instead exhibits oscillatory behaviors. Our method does not distinguish whether this is due to the onset of the $\mathbb{Z}_4$-ordered phase or a gapless incommensurate phase.
\pargap
\noindent\textbf{Correlation length extraction and scaling.} From the fluorescence pictures obtained at the end of an experimental sequence, we calculate the two-dimensional Rydberg density-density correlation map:

\begin{equation}
G(i,j) = \langle{n_{i}n_{j}}\rangle - \langle{n_i}\rangle\langle{n_j}\rangle.
\end{equation}

\begin{figure*}[htb]
  \centering
  \includegraphics[width=0.5\linewidth]{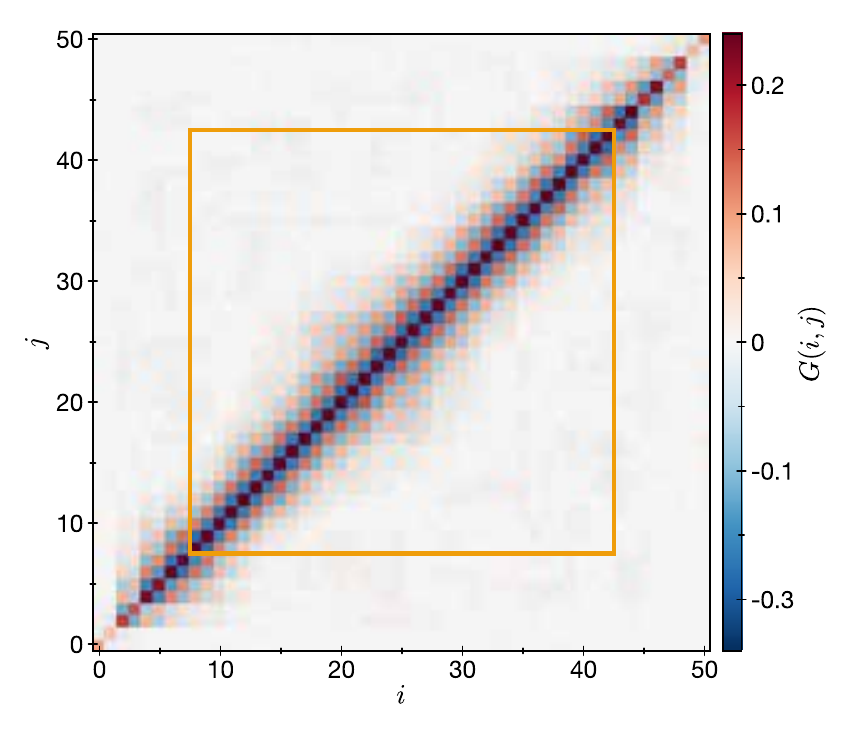}
  \caption{\textbf{Rydberg density-density correlations.} Full density-density correlation map for sites $i$ and $j$ after a slow sweep into the $\mathbb{Z}_2$-ordered phase. The orange square outline marks the bulk region used for analysis.}
  \label{sfig:g2_map}
\end{figure*}

To minimize boundary effects, we disregard 8 sites from each edge. From the remaining bulk correlations, we average out this map over diagonal lines of constant $|i-j|$ to obtain the Rydberg density-density correlation described in Eq. (2) in the main text (\ref{sfig:g2_map}). The uncertainties for the values of $G(r)$ are found via a jackknife analysis.

Two different approaches are used to extract a characteristic length from such correlations. For transitions into $\mathbb{Z}_N$-ordered states (Fig. 4), we  perform a least squares fit to the data with the model function:
\begin{equation}
    \label{eq:IndividualCorrelation}
    \hat{G}(r) = Ae^{-r/\xi}\hat{G}_{N}(r)_{\rm gs},
\end{equation}
where $A$ is the amplitude at $r=0$, $\xi$ is the correlation length, and $\hat{G}_{N}(r)_{\rm gs}$ is the ideal correlation function at integer values of $r$ for the corresponding $\mathbb{Z}_N$-ordered state, with a peak every $N$ sites: 
\begin{eqnarray}
  \hat{G}_{2}(r)_{\rm gs} &=&  \cos(2\pi r/2)  \nonumber \\
  \hat{G}_{3}(r)_{\rm gs} &=&  \cos(2\pi r/3)\\
  \hat{G}_{4}(r)_{\rm gs} &=&  \cos(2\pi r/4)+1/2\cos(2\pi r/2) \nonumber .
\end{eqnarray}
The range of distances used for all fits is $0<r\leq20$, where the cutoff at 20 sites is used to avoid any potential finite-size effects of the system. 

In addition to the procedure described above, for $\mathbb{Z}_2$-ordered states it is possible to extract a correlation length by fitting an exponential decay to the modulus of the correlation function, as is done in Fig. 2 of the main text. This method allows for  the determination of the correlation length in a way that is less susceptible to systematic effects arising from inversions of the alternating pattern, as observed in Fig. 3a of the main text. However, this method cannot be applied to $\mathbb{Z}_N$-ordered states for $N>2$, necessitating the use of a more general approach, such as the function $\hat{G}(r)$ defined above. While the scaling exponents extracted using both of these methods for the $\mathbb{Z}_2$-ordered state data are consistent within error bars, $\hat{G}(r)$ is used to obtain all exponents in Fig. 4c of the main text. 

\begin{figure*}[htb]
  \centering
  \includegraphics[width=0.66\linewidth]{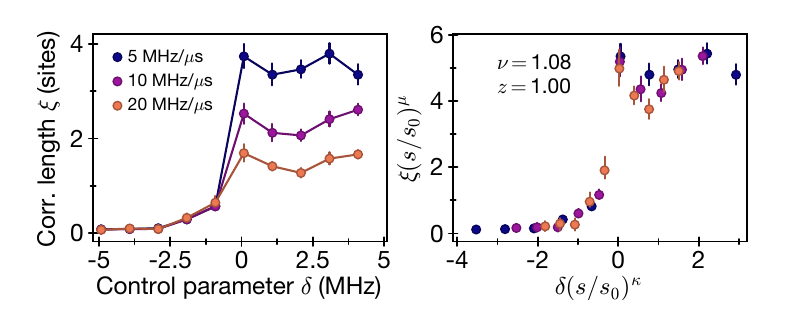}
  \caption{\textbf{Finite-size scaling across QPT into $\mathbb{Z}_2$-ordered phase.} \textbf{a,} Experimentally measured growth of the correlation length across the phase transition for different sweep speeds. The error bars denote the uncertainty of the power-law fit. \textbf{b,} Verification of critical exponents across the QPT into $\mathbb{Z}_{2}$-ordered phase by rescaling the control parameter and spatial correlations. Using the experimentally extracted value of the QKZM length scaling exponent, $\mu = 0.52$, and setting the dynamical critical exponent to the Ising prediction, $z=1$, it is observed that the data in \textbf{a} falls along a smooth function.}
  \label{sfig:Z2_time}
\end{figure*}

To extract the most likely scaling exponent $\mu$ at a given interaction, we fit a power law
\begin{equation}
    \xi = \xi_0\,(s/s_0)^{-\mu},
\end{equation}
where $s$ is the detuning sweep rate.
\pargap
\noindent\textbf{$\mathbb{Z}_N$ domain density.} In the fluorescence images obtained at the end of each experimental sequence, we identify the loss of an atom to a Rydberg excitation. In this way, we can directly count the number of instances of two lost atoms separated by $N$ sites, where every site in between contains an atom. To extract the data for Fig.~4b in the main text, we disregard the first 8 sites from the edges and count the instances in which both ends of the $N$ atom chain are within the bulk, $f_{N}$. The relative probability for two lost atoms separated by $N$ sites is given by:
\begin{equation}
	p_N = \frac{N\times f_{N}}{\sum_{i>0}(i\times f_{i})}.
\end{equation}
Unlike $G(r)$, $p_{N}$ is susceptible to detection infidelity~\cite{Bernien2017,Levine2018}.
\pargap
\noindent\textbf{Length rescaling of correlation functions.} In Fig.~3 of the main text, we use the normalized measured density-density correlation functions, $\frac{1}{A_{i}}G(r)_{i}$, and rescale the length $r$ by the QKZM length scaling exponent found via the scaling analysis of correlation length, $r~\rightarrow~(s/s_{0})^{\mu}r$.\pargap
\noindent\textbf{Finite-time scaling.} The length scaling exponent, $\mu$, found experimentally sets constraints to the possible combinations of the critical exponents $z$ and $\nu$ at a given interaction strength. In order to estimate, or qualitatively test, the possible values of $z$ and $\nu$, given the constraints set by $\mu$, we make use of the fact that in the critical region, all system properties scale in a universal way. The QKZM predicts a universal scaling of time with a scaling exponent of $z\nu/(1+z\nu)$, in addition to the scaling of length with $\mu = \nu/(1+z\nu)$~\cite{Gerster2018}
. In the experiment, the control parameter used to cross the quantum phase transition is $\delta = \Delta - \Delta_{c}$, where $\Delta_{c}$ is the value of the detuning at the critical point and can be estimated through numerical simulations (see section on numerical computation of the phase diagram). Near the critical point, the control parameter varies in time as $\delta(t) = st$, leading to a universal scaling of $\delta(s) = \delta_0(s_0/s)^{\kappa}$, where $\kappa = -1/(1+z\nu)$. Using the data shown in the main text for the correlation length growth across the transition into the $\mathbb{Z}_{2}$-ordered phase, we can apply the transformation $\xi\rightarrow\xi(s/s_{0})^{\mu}$ and $\delta \rightarrow \delta(s/s_{0})^{\kappa}$, to observe how well the data collapse to a universal shape. \ref{sfig:Z2_time} shows that these data are consistent with having critical exponents $z=1\simeq\nu$, as is expected for the Ising universality class. 
\pargap
\noindent\textbf{Numerical simulation of Kibble-Zurek dynamics.} We numerically model the dynamics of the system using matrix product states and employ a variant of a time evolving block decimation (TEBD) algorithm to propagate the state. We use a state update that allows us to exactly include the effect of interaction between atoms that are separated by less than $\ell=7$ sites. Interactions beyond this range are neglected. To this end, we use a Trotterization for the unitary that propagates the system from a time $t_k$ to a time $t_{k+1}=t_k+\Delta t$ as
\begin{align}
U(t_k\rightarrow t_{k+1})\approx\prod_{p=1}^{N-\ell} \exp\left(-i h_p(t_k) \Delta t\right),
\end{align}
where 
\begin{equation}
\begin{aligned}
  h_p(t_k) &=&\frac{1}{\ell}\sum_{j=0}^{\ell-1}\left(\frac{\Omega(t_k)}{2}\sigma_{p+j}^x-\Delta(t_k)n_{p+j}\right) \\
		   &+&\sum_{i=0}^{\ell-2}\sum_{j=i+1}^{\ell-1}\frac{1}{\ell-(j-i)}V_{i,j}n_{p+i} n_{p+j} \\
\end{aligned}
\end{equation}
for $1<p<N-\ell$, and $h_1$ and $h_{N-\ell}$ are similar, but with appropriately adjusted coefficients. 

We simulate the evolution according to the same pulse shape as applied in the experiment, with a time step of $\Delta t=0.15$~ns and a bond dimension of $128$. A comparison between the numerically simulated dynamics and the experimental results for different interaction strengths is shown in Fig.~4 of the main text. As described in the section on ``Correlation length extraction and scaling", deviations of the individual correlation functions from an exponentially decaying period-N density wave lead to systematic effects that dominate the uncertainty in the determination of the values presented in Fig.~4~b of the main text. The comparison between experimental and numerical results is susceptible to multiple effects, including finite-size effects~\cite{Jaschke2017}, accuracy of the approximate numerical methods used, experimental imperfections, and data fitting, which contribute to the observed discrepancy.
\pargap
\noindent\textbf{Chiral clock models.} QPTs in the Rydberg Hamiltonian, Eq. (1) of the main text, involving $\mathbb{Z}_n$ ($n \geq 3$) translational symmetry breaking along one spatial direction are  expected to be in the universality class of the extensively-studied $\mathbb{Z}_n$ {\it chiral} clock models~\cite{Huse1981,Ostlund1981,Huse1982,Huse1983,haldane1983phase,fendley2012parafermionic,Zhuang2015}. To elucidate this connection, let us focus on $n=3$ and consider the case when $V_1\gg \lvert \Omega\rvert,\lvert\Delta\rvert$, that is, nearest-neighbor interactions are strong enough to effectively preclude two neighboring atoms from simultaneously being in the Rydberg state. Since the van der Waals interactions decay rapidly as $V_x =C_6/x^6$, we neglect couplings beyond the third-nearest neighbor by approximating $V_x\approx 0$ for $x\geq 3$, leading to a truncated model of the form: 
\begin{align}\label{eq:Rydberg_toy}
H_{\rm Ryd}=&\sum_{i=1}^N \frac{\Omega}{2} (\ket{g_i}\!\bra{r_i}+\ket{r_i}\!\bra{g_i})-\Delta n_i +V_{2}n_i n_{i+2},
\end{align}
supplemented with the constraint $n_i\,n_{i+1}=0$.

The Hamiltonian \eqref{eq:Rydberg_toy} can be mapped to a system of hard-core bosons, where no more than one boson can occupy a single site. This follows upon identifying the state where the atom at site $i$ is in the internal state $\ket{r}$ ($\ket{g}$) with the presence (absence) of a boson. Defining the bosonic annihilation and number operators, $b_i$ and $n_i=b_i^\dag b_i$, respectively, we obtain
\begin{align}
\label{eq:UV}
H_b=\sum_{i=1}^{N} \frac{\Omega}{2}\,(b_i^\dag+b_i) - \Delta n_i +V_2\, n_i n_{i+2},
\end{align}
together with $n_i \, n_{i+1}=0$. This model (often referred to as the $U-V$ model) was shown by Refs.~\cite{Sachdev2002, Fendley2004} to exhibit a phase transition in the universality class of the three-state chiral clock model (CCM), over a set of parameters. 

The $\mathbb{Z}_n$ CCM is a simple extension of the transverse-field Ising model in which each spin is promoted to have $n>2$ states. However, instead of extending the symmetry from $\mathbb{Z}_2$ to $\mathbb{S}_n$, which would result in the $n$-state Potts model~\cite{wu1982potts}, the interactions are constructed to be invariant under $\mathbb{Z}_n$ transformations. With $n=3$, the three-state CCM is defined by the Hamiltonian~\cite{fendley2012parafermionic,Zhuang2015}
\begin{equation}
\label{eq:Hamiltonian}
H_{\textsc{ccm}} = -f \,\sum_{j=1}^N \tau_j^\dagger \,\mathrm{e}^{-\mathrm{i}\,\phi} - J \sum_{j =1}^{N-1} \sigma_j^\dagger \,\sigma_{j+1}\,\mathrm{e}^{-\mathrm{i}\,\theta} + \mbox{h.c.}
\end{equation}
acting on a one-dimensional chain of $N$ spins. The three-state spin operators $\tau_i$ and $\sigma_i$, which can be represented as
\begin{equation}
\tau=
  \begin{pmatrix}
    1 & 0 & 0 \\
    0 & \omega & 0\\
    0 & 0 & \omega^2
  \end{pmatrix},
  \quad
\sigma=
  \begin{pmatrix}
    0 & 1 & 0 \\
    0 & 0 & 1\\
    1 & 0 & 0
  \end{pmatrix} \,,
\end{equation}
act locally on the site $i$, and each satisfy 
\begin{align}
&\tau^3 = \sigma^3 = \II \,,
&\sigma\,\tau = \omega\,\tau\,\sigma\,; \quad \omega \equiv \exp\,(2\,\pi\,\mathrm{i}/3)\,.
\end{align}
Here, $\phi$ and $\theta$ define two chiral interaction phases: for describing spatially ordered phases, we need $\phi=0$, whereupon time-reversal and spatial-parity are both symmetries of the Hamiltonian but a purely spatial chirality is still present. Note that $H_{\rm Ryd}$ does not break time-reversal symmetry, necessitating the choice of $\phi=0$ in the quantum clock model \eqref{eq:Hamiltonian}. However, with both $\phi$ and $\theta$ nonzero, time-reversal and spatial-parity (inversion) symmetries are individually broken, but their product is preserved. 

As depicted in Fig.~4a, a generic state in the Hilbert space of the $\mathbb{Z}_3$ CCM can be mapped to one of three states of a clock according to the eigenvalue $1, \omega$, or $\omega^2$ of the operator $\sigma$ at each site. Consequently, there can be two domain walls in the system that differ in their energies, depending upon whether the clock rotates clockwise or counterclockwise upon crossing the wall. With $\phi = 0$ and $\theta \ne 0$, these have different energies, $2 J \sin(\pi/6 - \theta)$ and $2 J \sin(\pi/6 + \theta)$, and are thus inequivalent, leading to a chirality in the system that is absent for $\phi = \theta = 0$. 

On setting both $\phi = \theta = 0$,  $H_{\textsc{ccm}}$ reduces to the Hamiltonian for the three-state Potts model which possesses a larger symmetry, $\mathbb{S}_3$; the concomitant order--disorder phase transition has critical exponents $z = 1$, $\nu = 5/6$~\cite{alexander1975lattice,baxter1980hard, wu1982potts}, and accordingly $\mu \approx 0.45$. Note that these exponents are fundamentally distinct from those of the $\mathbb{Z}_3$ CCM, namely, $z \approx 1.33$, $\nu \approx 0.71$, yielding $\mu \approx 0.37$. The Rydberg Hamiltonian described in the main text contains a point along the phase boundary for which the condition of $\phi = \theta = 0$ is fulfilled, and with fine tuned pulses it may be possible to explore the critical properties of the three-state Potts model.

For $n=4$, the transitions of both the Potts and the achiral clock model are in the Ashkin-Teller universality class~\cite{jose1977renormalization, kadanoff1977connections}. The critical exponents of the four-state Potts model are $z = 1$, $\nu = 2/3$ (implying $\mu = 0.40$), whereas the four-state \textit{achiral} clock model is equivalent to two uncoupled Ising systems with $z = 1$, $\nu = 1$. With a nonzero chirality, however, it is believed that there is no direct transition from the ordered to the disordered phase in the four-state CCM as an intermediate gapless incommensurate phase always intervenes~\cite{Ostlund1981,haldane1983phase, yeomans1984annni}.

\FloatBarrier
\renewcommand{\addcontentsline}[3]{}

\bibliographystyle{naturemag}

\begin{thebibliography}{}

\bibitem[1]{Sachdev2009} Sachdev, S.  \emph{Quantum Phase Transitions} ({Cambridge University Press}, 2009), 2nd edn.

\bibitem[2]{Kibble1976} Kibble, T. W.~B. Topology of cosmic domains and strings. \emph{J. Phys. A: Math. Gen.} \textbf{9}, 1387 (1976).

\bibitem[3]{Zurek1985} Zurek, W.~H. Cosmological experiments in superfluid helium? \emph{Nature} \textbf{317}, 505 (1985).

\bibitem[4]{delCampo2014} del Campo, A. \& Zurek, W.~H. Universality of phase transition dynamics: Topological defects from symmetry breaking. \emph{International Journal of Modern Physics A} \textbf{29}, 1430018 (2014).

\bibitem[5]{Navon2015} Navon, N., Gaunt, A.~L., Smith, R.~P. \& Hadzibabic, Z. {Critical dynamics of spontaneous symmetry breaking in a homogenous Bose gas}. \emph{Science} \textbf{347}, 167 (2015).

\bibitem[6]{Polkovnikov2011} Polkovnikov, A., Sengupta, K., Silva, A. \& Vengalattore, M. Colloquium: Nonequilibrium dynamics of closed interacting quantum systems. \emph{Reviews of Modern Physics} \textbf{83}, 863 (2011).

\bibitem[7]{Polkovnikov2005} Polkovnikov, A. Universal adiabatic dynamics in the vicinity of a quantum critical point. \emph{Phys. Rev. B} \textbf{72}, 161201 (2005).

\bibitem[8]{Zurek2005} Zurek, W.~H., Dorner, U. \& Zoller, P. Dynamics of a quantum phase transition. \emph{Phys. Rev. Lett.} \textbf{95}, 105701 (2005).

\bibitem[9]{Dziarmaga2005} Dziarmaga, J. Dynamics of a quantum phase transition: Exact solution of the quantum ising model. \emph{Physical Review Letters} \textbf{95}, 245701 (2005).

\bibitem[10]{Huse1982} Huse, D.~A. \& Fisher, M.~E. {Domain Walls and the Melting of Commensurate Surface Phases}. \emph{Phys Rev Lett} \textbf{49}, 793 (1982).

\bibitem[11]{Ostlund1981} Ostlund, S. Incommensurate and commensurate phases in asymmetric clock models. \emph{Phys. Rev. B} \textbf{24}, 398 (1981).

\bibitem[12]{Tagliacozzo2013} Tagliacozzo, L., Celi, A., Orland, P., Mitchel, M.~W. \& Lewenstein, M. Simulation of non-abelian gauge theories with optical lattices. \emph{Nat. Commun.} \textbf{4} (2013).

\bibitem[13]{Weimer2010_b} Weimer, H., M{\"u}ller, M., Lesanovsky, I., Zoller, P. \& B{\"u}chler, H.~P. A {{Rydberg}} quantum simulator. \emph{Nat. Phys.} \textbf{6}, 382--388 (2010).

\bibitem[14]{Farhi2000} Farhi, E., Goldstone, J., Gutmann, S. \& Spiser, M. Quantum computation by adiabatic evolution. \emph{arXiv:quant-ph/0001106}.

\bibitem[15]{Gardas2018} Gardas, B., Dziarmaga, J., Zurek, W.~H. \& Zwolak, M. Defects in quantum computers. \emph{Sci. Rep.} \textbf{8} (2018).

\bibitem[16]{Anquez2016} Anquez, M. et~al. Quantum {K}ibble-{Z}urek mechanism in a spin-1 {B}ose-{E}instein condensate. \emph{Phys. Rev. Lett.} \textbf{116}, 155301 (2016).

\bibitem[17]{Clark2016} Clark, L.~W., Feng, L. \& Chin, C. Universal space-time scaling symmetry in the dynamics of bosons across a quantum phase transition. \emph{Science} \textbf{354}, 606 (2016).

\bibitem[18]{Endres2012} Endres, M. et~al. The ‘{H}iggs’ amplitude mode at the two-dimensional superfluid/{M}ott insulator transition.  \emph{Nature} \textbf{487}, 454 (2012).

\bibitem[19]{Chen2011} Chen, D., White, M., Borries, C. \& deMarco, B. Quantum quench of an atomic {M}ott insulator. \emph{Phys. Rev. Lett.} \textbf{106}, 235304 (2011).

\bibitem[20]{Braun2015} Braun, S. et~al. Emergence of coherence and the dynamics of quantum phase transitions. \emph{PNAS} \textbf{112}, 3641 (2015).

\bibitem[21]{Bernien2017} Bernien, H. et~al. Probing many-body dynamics on a 51-atom quantum simulator. \emph{Nature} \textbf{551}, 579 (2017).

\bibitem[22]{Samajdar2018} Samajdar, R., Choi, S., Pichler, H., Lukin, M.~D. \& Sachdev, S. Numerical study of the chiral $\mathbb{Z}_3$ quantum
  phase transition in one spatial dimension. \emph{Phys. Rev. A} \textbf{98}, 023614 (2018).

\bibitem[23]{Whitsitt2018} Whitsitt, S., Samajdar, R. \& Sachdev, S. Quantum field theory for the chiral clock transition in one spatial dimension. \emph{Phys. Rev. B} \textbf{98}, 205118 (2018).

\bibitem[24]{Chepiga2018} Chepiga, N. \& Mila, F. Floating phase versus chiral transition in a {1D} hard-boson model. \emph{Phys. Rev. Lett.} \textbf{122}  (2018).

\bibitem[25]{Kolodrubetz2012} Kolodrubetz, M., Clark, B.~K. \& Huse, D.~A. Nonequilibrium dynamical critical scaling of the
  quantum {I}sing chain. \emph{Phys. Rev. Lett.} \textbf{109} (2012).

\bibitem[26]{Cherng2006} Cherng, R.~W. \& Levitov, L.~S. Entropy and correlation functions of a driven quantum spin chain. \emph{Phys. Rev. A} \textbf{73}, 043614 (2006).

\bibitem[27]{Fendley2004} Fendley, P., Sengupta, K. \& Sachdev, S. Competing density-wave orders in a one-dimensional
  hard-boson model. \emph{Phys. Rev. B} \textbf{69}, 075106 (2004).

\bibitem[28]{haldane1983phase} Haldane, F. D.~M., Bak, P. \& Bohr, T. {Phase diagrams of surface structures from Bethe-ansatz solutions of the quantum sine-Gordon model}. \emph{Phys. Rev. B} \textbf{28}, 2743 (1983).

\bibitem[29]{Dutta2015} Dutta, A. \emph{et~al.} \emph{{Quantum phase transitions in transverse field spin models : From Statistical Physics to Quantum Information}} ({Cambridge University Press}, 2015).

\bibitem[30]{Moessner2001} Moessner, R., Sondhi, S.~L. \& Fradkin, E. Short-ranged resonating valence bond physics, quantum dimer models, and ising gauge theories. \emph{Phys. Rev. B} \textbf{65}, 024504 (2001).

\bibitem[31]{Barredo2018} Barredo, D., Lienhard, V., {de L{\'e}s{\'e}leuc}, S., Lahaye, T. \& Browaeys, A. Synthetic three-dimensional atomic structures assembled atom by atom. \emph{Nature} \textbf{561}, 79 (2018).

\bibitem[32]{Kumar2018} Kumar, A., Wu, T.-Y., Giraldo~Mejia, F. \& Weiss, D.~S. Sorting ultracold atoms in a three-dimensional optical lattice in a realization of maxwell's demon. \emph{Nature} \textbf{561}, 83 (2018).

\bibitem[33]{Pichler2018} Pichler, H., Wang, S.-T., Zhou, L., Choi, S. \& Lukin, M.~D. {Quantum Optimization for Maximum Independent Set
  Using Rydberg Atom Arrays}. \emph{arXiv:1808.10816}.

\end{thebibliography}

\begin{thebibliography}{}
\bibitem[34]{Endres2016} Endres, M. \emph{et~al.} Atom-by-atom assembly of defect-free one-dimensional cold atom arrays. \emph{Science} \textbf{354}, 1024 (2016).

\bibitem[35]{Levine2018} Levine, H. \emph{et~al.} High-fidelity control and entanglement of Rydberg-atom qubits. \emph{Phys. Rev. Lett.} \textbf{121}, 123603 (2018).

\bibitem[36]{white1992density} White, S. R. Density matrix formulation for quantum renormalization groups. \emph{Phys. Rev. Lett.} \textbf{69}, 2863 (1992).

\bibitem[37]{white1993density} White, S. R. Density-matrix algorithms for quantum renormalization groups. \emph{Phys. Rev. B} \textbf{48}, 10345 (1993).

\bibitem[38]{ostlund1995thermodynamic} Ostlund, S. \& Rommer, S. Thermodynamic limit of density matrix renormalization. \emph{Phys. Rev. Lett.} \textbf{75}, 3537 (1995).

\bibitem[39]{rommer1997class} Rommer, S. \& Ostlund, S. Class of ansatz wave functions for one-dimensional spin systems and their relation to the density matrix renormalization group. \emph{Phys. Rev. B} \textbf{55}, 2164 (1997).

\bibitem[40]{dukelsky1998equivalence} Dukelsky, J., Martin-Delgado, M. A., Nishino, T. \& Sierra, G. Equivalence of the variational matrix product method and the density matrix renormalization group applied to spin chains. \emph{EPL} \textbf{43}, 457 (1998).

\bibitem[41]{peschel1999density} Peschel, I., Wang, X., Kaulke, M. \& Hallberg, K. (eds.)\emph{Density-matrix Renormalization: A New Numerical Method in Physics.} Lecture Notes in Physics (Springer-Verlag, Berlin Heidelberg, 1999). 

\bibitem[42]{McCulloch2008} McCulloch, I. P. Infinite size density matrix renormalization group, revisited. \emph{arXiv:0804.2509v1} (2008).

\bibitem[43]{Pirvu2010} Pirvu, B., Murg, V., Cirac, J. I. \& Verstraete, F. Matrix product operator representations. \emph{New Journal of Physics} \textbf{12} (2010).

\bibitem[44]{schollwock2005density}Schollwock, U. The density-matrix renormalization group. \emph{Rev. Mod. Phys.} \textbf{77}, 259 (2005).

\bibitem[45]{schollwock2011density} Schollwock, U. The density-matrix renormalization group: a short introduction. \emph{Phil. Trans. R. Soc. A} \textbf{369}, 2643 (2011).

\bibitem[46]{Gerster2018} Gerster, M., Haggenmiller, B., Tschirsich, F., Silvi, P. \& Montangero, S. Dynamical Ginzburg criterion for the quantum-classical crossover of the Kibble-Zurek mechanism. \emph{arXiv:1807.10611} (2018).

\bibitem[47]{Jaschke2017} Jaschke, D., Maeda, K., Whalen, J.~D., Wall, M.~L. \& Carr, L.~D. Critical phenomena and {K}ibble-{Z}urek scaling in the long-range quantum {I}sing chain. \emph{New. J. Phys.} \textbf{19} (2017).

\bibitem[48]{Huse1981} Huse, D. A. Simple three-state model with infinitely many phases. \emph{Phys. Rev. B} \textbf{24}, 5180 (1981).

\bibitem[49]{Huse1983} Huse, D. A., Szpilka, A. M. \& Fisher, M. E. Melting and wetting transitions in the three-state chiral clock model. \emph{Physica A} \textbf{121}, 363 (1983).

\bibitem[50]{fendley2012parafermionic} Fendley, P. Parafermionic edge zero modes in $\mathbb{Z}_n$-invariant spin chains. \emph{J. Stat. Mech} \textbf{11}, 11020 (2012).

\bibitem[51]{Zhuang2015} Zhuang, Y., Changlani, H. J., Tubman, N. M. \& Hughes, T. L. Phase diagram of the Z3 parafermionic chain with chiral interactions. \emph{Phys. Rev. B} \textbf{92}, 035154 (2015).

\bibitem[52]{Sachdev2002} Sachdev, S., Sengupta, K. \& Girvin, S. M. Mott insulators in strong electric fields. \emph{Phys. Rev. B} \textbf{66}, 075128 (2002).

\bibitem[53]{wu1982potts} Wu, F.-Y. The Potts model. \emph{Rev. Mod. Phys.} \textbf{54}, 235 (1982).

\bibitem[54]{alexander1975lattice} Alexander, S. Lattice gas transition of He on Grafoil. A continous transition with cubic terms. \emph{Phys. Lett. A} \textbf{54}, 353 (1975).

\bibitem[55]{baxter1980hard} Baxter, R. J. Hard hexagons: exact solution. \emph{J. Phys. A: Math. Gen} \textbf{13}, L61 (1980).

\bibitem[56]{jose1977renormalization} Jos\'e, J. V., Kadanoff, L. P., Kirkpatrick, S. \& Nelson, D. R. Renormalization vortices, and symmetry-breaking perturbations in the two-dimensional planar model. \emph{Phys. Rev. B} \textbf{16}, 1217 (1977).

\bibitem[57]{kadanoff1977connections} Kadanoff, L. P. Connections between the critical behavior of the planar model and that of the eight-vertex model. \emph{Phys. Rev. Lett.} \textbf{39}, 903 (1977).

\bibitem[58]{yeomans1984annni} Yeomans, J. ANNNI and clock models. \emph{Physica B+C} \textbf{127}, 187 (1984).

\end{thebibliography}

\begin{table*}[htb]
\centering
 \includegraphics[width=0.5\linewidth]{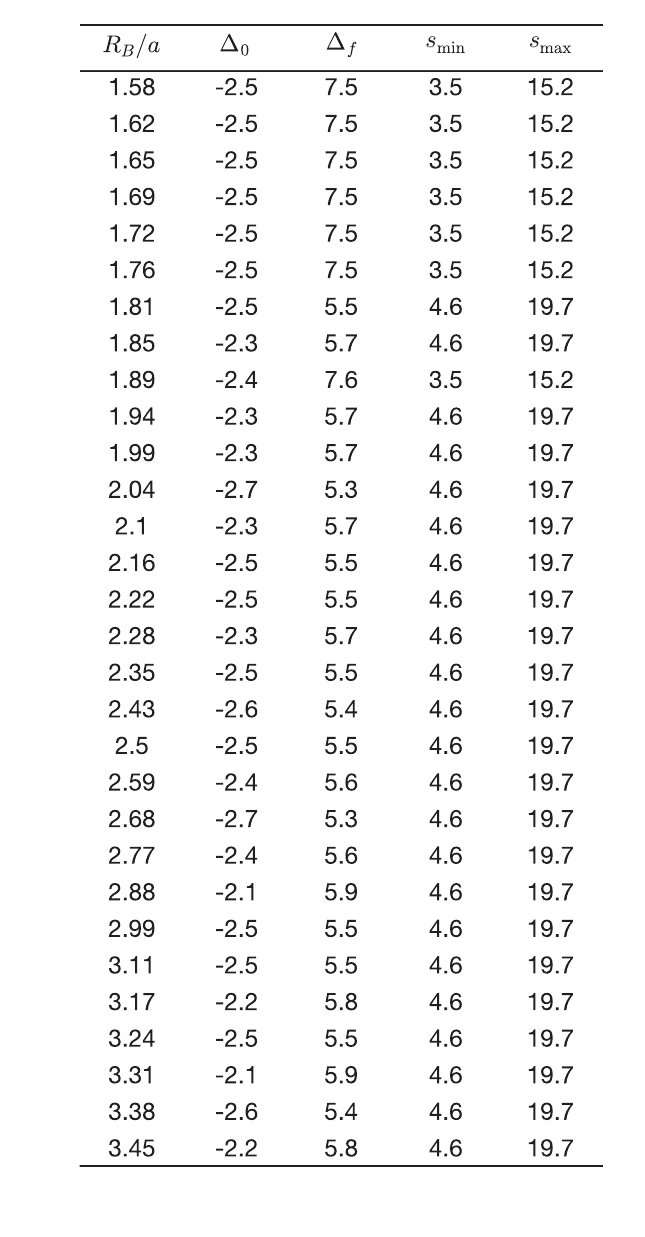}
\caption{\textbf{Pulse parameters for QKZM sweeps.} For different blockade radii $R_B/a$, we list the initial and final detunings $\Delta_0$ and $\Delta_f$ of the sweeps, and the minimum and maximum sweep speeds, $s_{\rm min}$ and $s_{\rm max}$, applied.
}
\label{table:exp_parameters}
\end{table*}

\end{document}